\documentclass[conference]{IEEEtran}
\IEEEoverridecommandlockouts
\usepackage{amsmath,amssymb,amsfonts}
\usepackage{graphicx}
\usepackage{bm}
\usepackage{booktabs}
\usepackage{url}
\usepackage{cite}
\usepackage{tikz}
\usetikzlibrary{arrows.meta,positioning,fit,backgrounds}

\newcommand{\E}{\mathbb{E}}
\newcommand{\C}{\mathbb{C}}
\newcommand{\herm}{\mathsf{H}}
\newcommand{\transp}{\mathsf{T}}
\newcommand{\norm}[1]{\left\lVert#1\right\rVert}

\begin{document}

\title{Koopman-Predictive Coefficient Tracking for\\ Vehicle-Mounted Monostatic ISAC}
\author{\IEEEauthorblockN{Anh Tuyen Le, Xiaojing Huang, and Y. Jay Guo}
\IEEEauthorblockA{Global Big Data Technologies Centre, University of
Technology Sydney, Australia\\
Email: \{anhtuyen.le, xiaojing.huang, jay.guo\}@uts.edu.au}}

\maketitle

\begin{abstract}
This paper studies predictive coefficient tracking for digitally generated self-interference cancellation (SIC) in vehicle-mounted monostatic integrated sensing and communication (ISAC) systems. In such systems, the canceller must suppress strong transmitter leakage and nearby unintended reflections while preserving delayed sensing echoes. On mobile platforms, mechanical vibration from engines, rotors, or nearby structures can induce quasi-periodic variation in the self-interference (SI) channel. The optimal cancellation coefficients may then vary within one adaptation period, causing the applied coefficients of a conventional adaptive filter to lag behind the moving SI channel and increasing the residual SI. To address this problem, we propose a Koopman-predictive SIC (KP-SIC) module that augments a conventional adaptive filter with a predicted coefficient trajectory. KP-SIC uses recent effective SI-channel coefficient snapshots to identify structured vibration-induced dynamics and predict a sample-varying coefficient sequence for the next adaptation period. The predicted coefficients place the canceller closer to the expected time-varying optimum, while the adaptive filter corrects residual prediction errors and unstructured variation. For evaluation, the module is integrated with a widely linear normalized least mean-square (WL-NLMS) filter under a flexible analog SIC (FASIC) framework. Simulations show that KP-SIC improves SI suppression by $6.3$--$10.2$~dB over conventional FASIC-WL-NLMS once the observation window spans at least about $1.5$ vibration cycles. Compared with a non-causal performance bound, KP-SIC achieves $60$--$74\%$ of the available prediction gain.
\end{abstract}


\section{Introduction}
\IEEEPARstart{M}{ONOSTATIC} integrated sensing and communication (ISAC) is attractive for mobile platforms that must communicate while sensing their surroundings, such as perceptive drones, connected vehicles, and autonomous robots. In-band full-duplex (IBFD) operation provides a natural way to support this capability, allowing the same transceiver to transmit probing or communication waveforms while receiving target echoes and uplink signals over the same time-frequency resource \cite{Liu22JSAC,Zhang21JSTSP,Tang24}. The main bottleneck is self-interference (SI), because leakage from the local transmitter can be far stronger than the target echo or uplink signal \cite{Sabharwal14,Smida24ProcIEEE}. For monostatic ISAC, cancellation must therefore suppress direct leakage and nearby unintended reflections while preserving delayed echoes that carry sensing information. This requirement calls for digitally controlled SIC structures that can shape the cancellation waveform through the coefficient vector and effective delay span, rather than only reducing the total transmit-correlated power.

In vehicle-mounted transceivers, the SI channel can also vary rapidly during operation. Rotating machinery, such as car engines or unmanned aerial vehicle (UAV) rotors, can vibrate the antennas and nearby reflecting structures, causing quasi-periodic changes in the SI propagation paths. These changes modulate the SI channel in a way similar to the micro-Doppler effect in radar returns \cite{Chen06,MorgeRollet22}. As a result, the effective SI-channel coefficients may contain a small number of structured spectral components at tens to hundreds of hertz, together with smaller unstructured variation \cite{Ruegg07,Bertocco25}. Conventional adaptive filters update the cancellation coefficients from past residual observations. Since convergence requires a finite number of samples, the applied coefficients can lag behind the time-varying SI channel when the channel changes within an adaptation period. This tracking lag increases the residual SI power even when the selected cancellation basis has sufficient delay span and accurately represents the SI path under static conditions.

To address this problem, this paper proposes a Koopman-predictive SIC (KP-SIC) module for digitally generated adaptive cancellation. The quasi-periodic structure of vibration-induced SI variation makes the future coefficient trajectory partly predictable from its recent evolution. Koopman operator methods are well suited to this setting because they can represent nonlinear dynamics through a linear model in an extended state space and can be identified directly from measured data \cite{Koopman31,Mezic05,Brunton22}, and they have recently been applied to prediction problems in wireless systems \cite{Krishnan25TVT,Girgis22}. KP-SIC uses recent estimates of the effective SI-channel coefficients, referred to as coefficient snapshots, to predict a sample-varying coefficient trajectory for the next adaptation period. The predicted trajectory places the applied coefficients closer to the expected time-varying optimum, while the adaptive filter corrects residual prediction errors and unstructured variations.

We demonstrate the effectiveness of KP-SIC by integrating it with flexible analog SIC (FASIC) \cite{Le25SPAWC,LeISACjournal}, where a digitally generated cancellation waveform is injected in the RF domain before the receiver analog-to-digital converter (ADC). In the resulting KP-SIC-assisted FASIC scheme, the Koopman branch predicts the cancellation-coefficient trajectory, and a widely linear normalized least mean square (WL-NLMS) filter corrects the remaining mismatch. The scheme is evaluated against conventional WL-NLMS tracking and a non-causal performance bound. Simulation results under structured SI-channel variation show that KP-SIC improves SI suppression by $6.3$--$10.2$~dB over conventional FASIC-WL-NLMS and recovers $60$--$74\%$ of the predictive gain identified by the non-causal bound. These gains are achieved once the observation window spans sufficient vibration cycles to identify the dominant modes, yielding a practical window-sizing condition for vibration-aware tracking.

The rest of this paper is organized as follows. Section~II introduces the signal model and formulates the coefficient-tracking problem. Section~III presents the proposed KP-SIC method. Section~IV evaluates the convergence, variation-rate dependence, and robustness to unstructured SI-channel variation through simulations. Finally, Section~V concludes the paper.
\section{Signal Model}\label{sec:model}
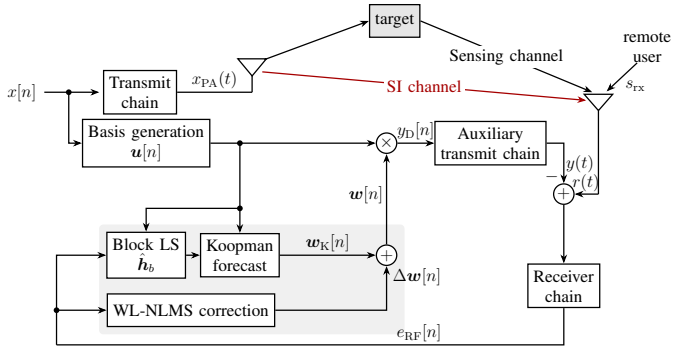
\begin{figure}[!t]\centering
\resizebox{\columnwidth}{!}{%
\begin{tikzpicture}[x=1cm,y=1cm,>={Stealth[length=2.2mm]},line width=0.9pt,
 font=\large,
 blk/.style={draw,fill=white,align=center,inner sep=4pt,minimum height=8mm},
 sm/.style={draw,circle,fill=white,inner sep=0pt,minimum size=5mm},
 lbl/.style={inner sep=2pt},
 dom/.style={font=\large\itshape,text=black!60}]

% ------------------------------------------------------- transmit path
\node[lbl] (xin) at (-0.15,6.6) {$x[n]$};
\draw[->] (0.4,6.6) -- (1.75,6.6);
\fill (1.0,6.6) circle (0.07);
\node[blk] (tc) at (2.75,6.6) {Transmit\\chain};
\draw[-] (tc.east) -- (5.5,6.6) node[midway,above,lbl] {$x_{\rm PA}(t)$};
\draw[-] (5.5,6.6) -- (5.5,7.0);
\draw[-] (5.15,7.42) -- (5.85,7.42) -- (5.5,7.0) -- cycle;

% ------------------------------------------------------- over the air
\draw[->,red!65!black] (5.75,7.1) -- (13.7,6.4)
  node[pos=0.5,fill=white,inner sep=2pt,text=red!65!black] {SI channel};
\node[blk,fill=black!10] (tg) at (9.0,8.35) {target};
\draw[->] (5.8,7.4) -- (tg.west);
\draw[->] (tg.east) -- (13.85,6.6)
  node[pos=0.5,fill=white,inner sep=2pt] {Sensing channel};
\node[lbl,align=center] (ue) at (15.2,7.7) {remote\\user};
\draw[->] (14.95,7.32) -- (14.2,6.62) node[midway,below right,lbl] {$s_{\rm rx}$};

% ------------------------------------------------------- receive path
\draw[-] (13.65,6.55) -- (14.35,6.55) -- (14.0,6.15) -- cycle;
\draw[-] (14.0,6.15) -- (14.0,4.1);
\node[sm] (cmb) at (13.15,4.1) {$+$};
\draw[->] (14.0,4.1) -- (cmb.east);
\node[lbl] at (13.7,4.38) {$r(t)$};
\node[lbl] at (12.86,4.52) {$-$};
\node[blk] (rc) at (13.15,1.85) {Receiver\\chain};
\draw[->] (cmb.south) -- (rc.north);

% ------------------------------------------------------- cancellation path
\node[blk] (bg) at (2.9,5.35) {Basis generation\\$\bm u[n]$};
\draw[->] (1.0,6.6) -- (1.0,5.35) -- (bg.west);
\node[sm] (mult) at (8.8,5.35) {$\times$};
\draw[->] (bg.east) -- (mult.west);
\fill (5.2,5.35) circle (0.07);
\node[blk] (atc) at (11.35,5.35) {Auxiliary\\transmit chain};
\draw[->] (mult.east) -- (atc.west)
  node[midway,above,lbl] {$y_{\rm D}[n]$};
\draw[->] (atc.east) -| (cmb.north);
\node[lbl] at (13.55,4.85) {$y(t)$};

% ------------------------------------------------------- tracker group
\node[blk] (bls) at (2.9,2.6) {Block LS\\$\hat{\bm h}_b$};
\node[blk] (kf) at (5.2,2.6) {Koopman\\forecast};
\node[blk] (wln) at (4.0,1.25) {WL-NLMS correction};
\node[sm] (wsum) at (8.8,2.6) {$+$};
\draw[->] (bls) -- (kf);
\draw[->] (kf.east) -- (wsum.west) node[midway,above,lbl] {$\bm w_{\rm K}[n]$};
\draw[->] (wln.east) -| (wsum.south)
  node[pos=0.88,right=2pt,lbl] {$\Delta\bm w[n]$};
\draw[->] (wsum.north) -- (mult.south)
  node[pos=0.55,left,lbl] {$\bm w[n]$};
\begin{scope}[on background layer]
\node[fill=black!6,rounded corners=3pt,inner sep=5pt,
      fit=(bls)(kf)(wln)(wsum)] (grp) {};
\end{scope}

% ------------------------------------------------------- feedback and taps
\draw[->] (5.2,5.35) -- (kf.north);
\fill (5.2,3.75) circle (0.07);
\draw[->] (5.2,3.75) -- (2.9,3.75) -- (bls.north);
\draw (rc.south) -- (13.15,0.4) -- (0.7,0.4);
\node[lbl,above] at (9.6,0.4) {$e_{\rm RF}[n]$};
\draw[->] (0.7,0.4) -- (0.7,2.6) -- (bls.west);
\fill (0.7,1.25) circle (0.07);
\draw[->] (0.7,1.25) -- (wln.west);
\end{tikzpicture}}
\caption{KP-SIC-assisted FASIC transceiver with the proposed coefficient
tracker (shaded). $x_{\rm PA}(t)$, $y(t)$, and $r(t)$ are RF waveforms.}
\vspace{-2.5mm}
\label{fig:arch}
\end{figure}

Let $x[n]$ denote the known discrete-time baseband transmit sample of the IBFD ISAC transceiver shown in Fig.~\ref{fig:arch}. The complex-envelope signal at the RF combiner input before cancellation is modeled as
\begin{equation}
r[n]=r_{\rm SI}[n]+s_{\rm rx}[n]+\eta[n], \label{eq:rx-signal}
\end{equation}
where $r_{\rm SI}[n]$ is the SI, $s_{\rm rx}[n]$ contains the useful target echoes and remote communication signals, and $\eta[n]$ is the receiver noise. The SI is caused by direct leakage and nearby reflection paths, whose complex-envelope contribution is modeled as \begin{equation}
r_{\rm SI}[n]=\sum_{l=1}^{L} g_l[n]\,x_{\rm PA}[n-\tau_l], \label{eq:channel}
\end{equation}
where $x_{\rm PA}[n]$ is the sampled complex envelope of the amplified transmit waveform, and path $l$ has delay $\tau_l$ and time-varying complex gain $g_l[n]$. The delay $\tau_l$ is expressed in symbol periods $T_s$ and is not necessarily aligned with the sampling grid. Fixed attenuation due to passive isolation, coupling loss, and path loss is absorbed into $g_l[n]$.

Mechanical motion perturbs the effective lengths of the SI propagation
paths. Even a sub-wavelength displacement of the antennas or nearby
reflectors can produce a noticeable carrier-phase shift in a leakage path,
following the same physical mechanism that gives rise to micro-Doppler
modulation in radar returns \cite{Chen06,MorgeRollet22}. We model this
vibration-induced variation through the time-varying path gain \cite{Long24}
\begin{equation}
g_l[n]=\bar g_l\exp\!\Big(\mathrm{j}\Big[\sum_{k=1}^{2} \theta_k\cos(2\pi \nu_k n+\phi_{l,k})+\sigma_v\,\varepsilon_l[n]\Big]\Big), \label{eq:gains}
\end{equation}
where $\bar g_l$ is the nominal complex gain. The two sinusoidal phase
terms represent structured quasi-periodic motion, with phase depths
$\theta_k$, normalized frequencies $\nu_k$ in cycles per sample, and
path-dependent phases $\phi_{l,k}$. Each depth $\theta_k$ is the peak
phase shift induced by the corresponding motion and is determined by the
displacement relative to the carrier wavelength. The additional term
$\sigma_v\varepsilon_l[n]$ captures unstructured variation, the part of
the motion with no periodic pattern. The process $\varepsilon_l[n]$ is
real with unit variance, so $\sigma_v$ sets the strength of this
variation in radians.

The SI component is approximated using a cancellation basis generated from
the known transmit samples. Let $\bm u[n]\in\C^M$ collect the corresponding
basis functions, which may include delayed, widely linear (WL), or nonlinear
terms depending on the transmitter impairments and SI coupling paths to be
represented. For a selected basis, the SI signal can be decomposed as
\begin{equation}
r_{\rm SI}[n]=\bm h[n]^{\transp}\bm u[n]+\xi[n],
\label{eq:equiv}
\end{equation}
where $\bm h[n]$ is the effective SI-channel coefficient vector and $\xi[n]$
is the projection error, which is orthogonal to the selected basis by
construction. Since the physical path gains in
\eqref{eq:gains} are time-varying, the effective coefficient vector
$\bm h[n]$ also contains the structured quasi-periodic and unstructured
variation of the SI channel. The delay span of the basis determines which
transmit-correlated components can be represented by the canceller. In
monostatic ISAC, this span should cover the direct leakage and nearby
unintended reflections, while avoiding unnecessary overlap with delayed
target echoes that carry sensing information. Remote communication signals
and receiver noise are assumed to be independent of the local transmit
basis. Target echoes are still related to the transmit waveform, but echoes
that are sufficiently separated in delay have negligible correlation with the
finite-span cancellation basis. The cancellation basis therefore satisfies
\[
\E\{\bm u^*[n]\xi[n]\}=\bm 0,\;
\E\{\bm u^*[n]s_{\rm rx}[n]\}=\bm 0,\;
\E\{\bm u^*[n]\eta[n]\}=\bm 0.
\]
Under these conditions, coefficient tracking suppresses the SI components
represented by the basis without forcing the useful receive signal to be
canceled.

A digitally controlled canceller synthesizes the baseband cancellation samples
from the selected basis as
\begin{equation}
y_{\rm D}[n]=\bm w[n]^{\transp}\bm u[n],
\label{eq:digital-synthesis}
\end{equation}
where $\bm w[n]$ is the adjustable cancellation coefficient vector. The
samples $y_{\rm D}[n]$ are converted through an auxiliary digital-to-analog
converter (DAC) and upconverter into the RF cancellation waveform $y(t)$,
which is combined destructively with the received signal before the receiver
ADC.
Assuming an ideal and calibrated auxiliary path, the complex-envelope signal
after RF cancellation is
\begin{equation}
e_{\rm RF}[n]=r[n]-y_{\rm D}[n].
\label{eq:rf-sum}
\end{equation}
Although \eqref{eq:rf-sum} is written in discrete time, it represents
cancellation performed in the RF domain before ADC sampling. Substituting
\eqref{eq:rx-signal}, \eqref{eq:equiv}, and \eqref{eq:digital-synthesis} into
\eqref{eq:rf-sum} gives
\begin{equation}
e_{\rm RF}[n]=\big(\bm h[n]-\bm w[n]\big)^{\transp}\bm u[n]
      +s_{\rm rx}[n]+\xi[n]+\eta[n].
\label{eq:tracking-residual}
\end{equation}
Under the orthogonality conditions above, the coefficient vector that
minimizes the SI-dependent contribution to the residual power is
$\bm w_{\rm opt}[n]=\bm h[n]$.

Conventional adaptive filtering algorithms approach this optimum by updating
$\bm w[n]$ from the measured residual. During continuous tracking, the
coefficient vector obtained in one adaptation period is used as the starting
point for the next, where an adaptation period denotes a fixed block of
samples, used throughout as the reference interval for coefficient
tracking. This reactive update works
well when the SI channel changes slowly, but it becomes less effective when
$\bm h[n]$ varies appreciably within an adaptation period. In that case, the
optimum $\bm w_{\rm opt}[n]$ moves before the adaptive filter converges, so
the applied coefficients lag behind the SI channel and the residual SI power
increases. Increasing the step size can shorten the response time, but it
also amplifies gradient noise and limits the tracking performance of a purely
reactive loop. A predicted coefficient trajectory can instead hold the applied
coefficients closer to the expected optimum throughout the adaptation
period, leaving the adaptive filter to correct only the remaining mismatch.
This motivates the KP-SIC tracking method developed in the next section.

\section{Koopman-Predictive SIC Tracking}\label{sec:method}

The proposed KP-SIC module, shown shaded in Fig.~\ref{fig:arch}, acts on the
coefficient vector of the digitally controlled canceller. The known transmit
samples generate the basis $\bm u[n]$, and the applied coefficients synthesize
the digital cancellation waveform in \eqref{eq:digital-synthesis}. KP-SIC
augments the adaptive canceller by writing the applied coefficients as
\begin{equation}
\bm w[n]=\bm w_{\rm K}[n]+\Delta\bm w[n],
\label{eq:split}
\end{equation}
where $\bm w_{\rm K}[n]$ is the Koopman-predicted component and
$\Delta\bm w[n]$ is the adaptive correction. The predicted component captures
structured SI-channel variation, while the adaptive correction compensates
for prediction errors and unstructured variation. The resulting
predictive-adaptive loop is referred to as the \emph{assisted} loop. When
$\bm w_{\rm K}[n]=\bm 0$, the conventional adaptive canceller is recovered.

To predict the SI-channel evolution, KP-SIC first forms a sequence of
effective coefficient snapshots. Although the receiver ADC observes the
signal after RF cancellation, the applied cancellation waveform is known.
Under the ideal and calibrated auxiliary-chain assumption, the equivalent
pre-cancellation observation is reconstructed as
\[
r[n]=e_{\rm RF}[n]+y_{\rm D}[n].
\]
During each adaptation period $b$, a block least-squares (LS) stage
estimates the effective SI-channel coefficient vector $\hat{\bm h}_b$ from
the reconstructed samples and the known basis $\bm u[n]$. The sequence
$\hat{\bm h}_b$ then provides snapshots of the SI-channel dynamics. The
Koopman forecast stage predicts a sample-varying coefficient trajectory
$\bm w_{\rm K}[n]$ for the next adaptation period from these snapshots.
Koopman theory represents nonlinear state evolution through a linear
operator acting on functions of the state
\cite{Koopman31,Mezic05,Brunton22}. The forecast supplies the structured
component of the coefficient update, while the WL-NLMS correction updates
$\Delta\bm w[n]$ at every sample from the measured residual.

\subsection{Forecast branch}
The forecast branch maps the recent coefficient snapshots
$\hat{\bm h}_b$ to a sample-varying coefficient trajectory for the next
adaptation period. Since the cancellation basis functions are generally
correlated, coefficient errors are first weighted by the Gram matrix
\begin{equation}
\bm G=\frac{1}{N_{\rm tot}}\sum_{n=1}^{N_{\rm tot}}
\bm u^*[n]\bm u^{\transp}[n],
\label{eq:gram}
\end{equation}
where $N_{\rm tot}$ is the length of the known transmit frame used to form
$\bm G$. Each coefficient snapshot is then transformed as
\begin{equation}
\bm q_b=\bm G^{1/2}\hat{\bm h}_b .
\label{eq:whiten}
\end{equation}
In this coordinate system, the squared Euclidean error corresponds to the
average residual SI power caused by the associated coefficient error.

To capture the temporal evolution of the effective SI channel, the $d$ most
recent transformed snapshots are arranged into the lifted state
\begin{equation}
\bm z_b=[\bm q_b^{\transp},\ldots,\bm q_{b-d+1}^{\transp},
\bm q_b^{\herm},\ldots,\bm q_{b-d+1}^{\herm}]^{\transp}.
\label{eq:lift}
\end{equation}
The delay depth $d$ provides the temporal memory of the forecast, while the
conjugate entries allow the lifted state to represent improper complex
variations arising in widely linear cancellation \cite{Picinbono95}.

A low-rank linear model is identified for the lifted-state evolution,
$\bm z_{b+1}\approx\bm A\bm z_b$, using dynamic mode decomposition (DMD)
\cite{Schmid10,Williams15}. From a window of $W$ snapshots, the $K=W-d+1$
lifted states form the shifted data matrices
\[
\bm Z_1=[\bm z_{b-K+1},\ldots,\bm z_{b-1}],\qquad
\bm Z_2=[\bm z_{b-K+2},\ldots,\bm z_b].
\]
With the economy-size singular value decomposition
$\bm Z_1=\bm U\bm\Sigma\bm V^{\herm}$, the reduced operator on the dominant
rank-$r$ subspace is
\begin{equation}
\tilde{\bm A}=\bm U_r^{\herm}\bm Z_2 \bm V_r\bm\Sigma_r^{-1}.
\label{eq:dmd}
\end{equation}
Let $\tilde{\bm A}\bm Y=\bm Y\bm\Lambda$ be its eigendecomposition. The DMD
modes in the lifted space are $\bm\Phi=\bm U_r\bm Y$. To avoid artificial
forecast growth, eigenvalues satisfying $|\lambda_i|>\rho$ are projected
onto the unit circle as $\lambda_i\leftarrow\lambda_i/|\lambda_i|$, where
$\rho$ is chosen slightly below one \cite{Choi24MLSP}.

The modal amplitudes $\bm a_b$ describe how strongly each DMD mode is
present in the recent lifted-state sequence. They are fitted jointly over
the sliding window, so the forecast is based on the recent evolution rather
than on a single noisy snapshot. The predicted lifted state at fractional
position $\tau\in(0,1]$ of the next adaptation period is
\begin{equation}
\hat{\bm z}_{b+1}(\tau)
=\bm\Phi\bm\Lambda^\tau\bm a_b .
\label{eq:predstate}
\end{equation}
Let $\bm E$ select the leading $M$ entries of the lifted state, which
correspond to the current transformed coefficient vector. Reversing the
Gram-matrix transformation gives the predicted effective SI-channel
coefficient vector
\begin{equation}
\hat{\bm h}_{b+1}(\tau)
=\bm G^{-1/2}\bm E\hat{\bm z}_{b+1}(\tau)
=\bm G^{-1/2}\bm E\bm\Phi\bm\Lambda^\tau\bm a_b .
\label{eq:forecast}
\end{equation}
Since the optimal cancellation coefficients equal the effective SI-channel
coefficients under the model in Section~\ref{sec:model}, KP-SIC sets
$\bm w_{\rm K}(\tau)=\hat{\bm h}_{b+1}(\tau)$. Evaluating this prediction at
the sample positions of the next adaptation period produces the applied
sample-wise sequence $\bm w_{\rm K}[n]$. The forecast is formed at the end
of adaptation period $b$ and applied over period $b+1$.

\subsection{FASIC integration with WL-NLMS}
KP-SIC can be combined with different adaptive filtering algorithms. For the
waveform-level demonstration, we integrate it with the FASIC realization in
Fig.~\ref{fig:arch}. The FASIC basis is a widely linear expanded memory
polynomial (EMP), whose delayed direct and conjugate branches represent
transmitter I/Q imbalance, power-amplifier (PA) distortion, and multipath SI
coupling \cite{Le25SPAWC}. The basis vector $\bm u[n]$ therefore stacks
delayed EMP terms of the transmit sample together with their conjugates. In
the assisted loop, the applied coefficient vector is given by
\eqref{eq:split}, while only the correction term $\Delta\bm w[n]$ is updated
adaptively. Substituting \eqref{eq:split} into \eqref{eq:tracking-residual}
replaces the coefficient error $\bm h[n]-\bm w[n]$ by
$\bm h[n]-\bm w_{\rm K}[n]-\Delta\bm w[n]$, where $\bm w_{\rm K}[n]$ is
supplied by the Koopman forecast. It varies
with $n$ across the adaptation period but is not adapted by the WL-NLMS
update. Therefore, the adaptive filter does
not need to estimate the full effective SI-channel coefficient vector
$\bm h[n]$. Instead, it only tracks the residual coefficient error
$\bm h[n]-\bm w_{\rm K}[n]$. The WL-NLMS correction is then updated from the
measured residual as
\begin{equation}
\Delta\bm w[n{+}1]=\Delta\bm w[n]+
\frac{\mu\bm u^*[n]e_{\rm RF}[n]}
{\epsilon+\norm{\bm u[n]}^2},
\label{eq:nlms}
\end{equation}
where $\mu$ is the step size and $\epsilon$ prevents division by a small
input power. Under the orthogonality conditions in
Section~\ref{sec:model}, the correction term that minimizes the
SI-dependent residual power is
\begin{equation}
\Delta\bm w_{\rm opt}[n]=\bm h[n]-\bm w_{\rm K}[n].
\label{eq:corr-opt}
\end{equation}
Thus, KP-SIC changes the target of adaptation rather than the WL-NLMS update
itself. Conventional FASIC adapts toward $\bm h[n]$, whereas the assisted
loop adapts toward the remaining error after prediction. When the Koopman
forecast captures the structured SI-channel variation, this residual target
varies more slowly than $\bm h[n]$, allowing WL-NLMS to correct the mismatch
with less tracking lag. The Koopman stage is executed once per adaptation
period, while the WL-NLMS correction remains the only sample-wise update.

\section{Simulation Results}\label{sec:results}
\subsection{Simulation setup}
The proposed KP-SIC-assisted FASIC scheme is evaluated through
waveform-level simulations and compared with conventional and predictive
FASIC tracking schemes. The transceiver operates at a carrier frequency of
$3.5$~GHz. The transmitted signal is a quadrature phase-shift keying (QPSK)
waveform with root-raised-cosine pulse shaping, roll-off factor $0.25$, and
four samples per symbol. The transmitter includes widely linear (WL) I/Q
gain and phase imbalance of $1$~dB and $-5^\circ$, respectively, followed by
a fifth-order PA nonlinearity. The receiver noise is set $60$~dB below the
received SI power, so its contribution to the reported post-cancellation
residual is negligible. The canceller uses a WL expanded memory polynomial
(EMP) basis with six basis functions and $12$ taps spaced at $T_s/2$. The
direct and conjugate branches give $M=144$ coefficients. Each adaptation
period spans $N=512$ waveform samples, which provides enough samples for a
well-conditioned block-LS snapshot of the $M$ coefficients. The WL-NLMS step
size is $\mu=1.0$. The Koopman stage uses a window length $W=48$, rank
$r=14$, embedding depth $d=6$, and eigenvalue threshold $\rho=0.9$. The
WL-NLMS loop is not restarted at period boundaries, but continues to update
at every sample throughout the run. Each simulation run contains $240$
adaptation periods, with the first $60$ periods excluded from steady-state
statistics. Unless otherwise stated, results are averaged over six
independent realizations, with four realizations used for parameter sweeps.

The SI channel in \eqref{eq:channel} contains three paths with delays
$[0,\,0.9,\,3.3]T_s$ and $-35$~dB passive isolation. The vibration-induced
phase terms in \eqref{eq:gains} use normalized frequencies $\nu_1=f_v/N$ and
$\nu_2=0.618\nu_1$, where $f_v$ is the primary motion rate in cycles per
adaptation period. The phase depths are $\theta_1=0.25$ and
$\theta_2=0.10$~rad, corresponding to path-length excursions of
approximately $3.4$~mm and $1.4$~mm at the carrier frequency. With a $1$~ms
adaptation period, the sweep $f_v=0.01$--$0.4$ corresponds to vibration rates from $10$ to $400$~Hz, which covers with margin the roughly $30$--$210$~Hz vibration lines reported for road vehicles and rotorcraft \cite{Ruegg07,Bertocco25}. The second tone uses the noninteger ratio $0.618$ to avoid an
artificially repeating two-tone pattern, while both tones remain resolvable
from the coefficient snapshots over the rate sweep. The unstructured
component in \eqref{eq:gains} follows the correlated-noise model of
\cite{Baddour05}. It varies smoothly over approximately two adaptation
periods, has nominal strength $\sigma_v=0.02$~rad, and is varied in the
unstructured-motion study.

SI suppression is defined as the ratio between the received SI power before
cancellation and the total residual power after the RF combiner. The
optimal coefficient vector is
obtained by projecting the noiseless SI waveform onto the EMP basis and is
used only for evaluation. Using this optimal coefficient sequence in place
of the forecast gives a non-causal performance bound, representing the ideal
case in which the SI-channel variation is known.

\subsection{Convergence behavior}
\begin{figure}[!t]\centering
\includegraphics[width=0.80\linewidth]{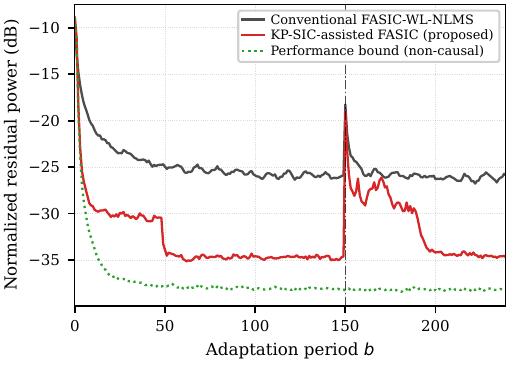}
\caption{Residual power after cancellation, normalized to the received
SI power, averaged over $50$ transmit-data realizations at $f_v=0.05$.
The vertical line marks the abrupt SI-channel change.}
\vspace{-2.5mm}
\label{fig:conv}
\end{figure}

Fig.~\ref{fig:conv} shows the residual power at the RF combiner output,
normalized to the received SI power and averaged over $50$ transmit-data
realizations at $f_v=0.05$. The assisted scheme reaches a deeper
cancellation level than conventional FASIC-WL-NLMS throughout the run.
During the first $W$ adaptation periods, when the forecast window is being
filled, the assisted loop already operates about $7$~dB below the
conventional loop. Once the Koopman forecast becomes active, the residual
decreases by a further $5$~dB and reaches $34.7$~dB below the received SI.
In comparison, the conventional loop reaches $25.7$~dB, while the
non-causal performance bound reaches $38.0$~dB.

The lower residual before the abrupt change shows that the
Koopman-predicted component reduces the coefficient error that must be
corrected by WL-NLMS. Even before the forecast is activated, the assisted
loop benefits from using the latest block-LS snapshot as the predictive
component. After the window is filled, the Koopman branch provides a
sample-varying prediction of the structured SI-channel motion, which
further reduces the residual gap to the non-causal performance bound.

To evaluate robustness to events outside the vibration model, an abrupt
channel change is introduced at adaptation period $150$. At this point,
every path gain in \eqref{eq:gains} is multiplied by an additional factor
with magnitude $0.3$ and random phase. This perturbation represents a
sudden environmental change, such as a new scatterer entering the near
field, and is not predictable from past snapshots. The post-change convergence time is defined as the number of periods required
for the suppression to return within $1$~dB of the common pre-change
conventional-FASIC level for three consecutive periods. This common
reference is used because the schemes have different steady-state
suppression levels.

After the abrupt change, both schemes initially lose cancellation to a
similar level. However, the assisted loop returns to the common pre-change
conventional level in $1.8\pm1.2$ adaptation periods, compared with
$6.7\pm2.4$ periods for the conventional loop. Throughout the transient, the assisted loop remains above the
conventional loop in every realization.

\subsection{Variation rate}
\begin{figure}[!tb]\centering
\includegraphics[width=0.80\linewidth]{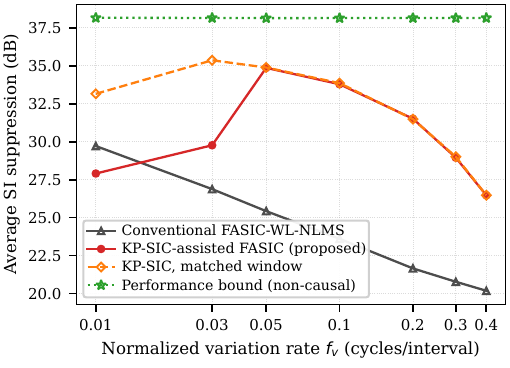}
\caption{Average SI suppression versus normalized SI variation rate.}
\vspace{-2.5mm}
\label{fig:rate}
\end{figure}

We next evaluate the tracking performance as the structured SI-channel
variation becomes slower or faster relative to the adaptation period.
Fig.~\ref{fig:rate} shows the average SI suppression as the normalized
variation rate $f_v$ increases. Conventional FASIC-WL-NLMS loses $9.5$~dB
over the sweep, decreasing from $29.7$~dB at $f_v=0.01$ to $20.2$~dB at
$f_v=0.4$. With KP-SIC, the assisted scheme improves suppression by
$6.3$--$10.2$~dB for $f_v\geq0.05$. The largest gain occurs at $f_v=0.1$,
where the SI channel changes appreciably within one adaptation period but
remains predictable from recent coefficient snapshots.

The suppression of the assisted scheme is therefore not monotonic in
$f_v$, and the shape of the curve is set by the observation window rather
than by the tracking rate. A window of $W$ adaptation periods spans $Wf_v$
vibration cycles. At $f_v=0.01$ it covers only $0.5$ of a cycle, so the
dominant modes cannot be identified and the forecast is less accurate than
simply holding the latest snapshot, leaving the suppression $1.8$~dB below
conventional FASIC. At $f_v=0.05$ the same window spans $2.4$ cycles, the
modes are resolved, and the prediction gain appears.

This behavior gives a practical window-sizing condition: the observation
window should span at least about $1.5$ vibration cycles, that is,
$Wf_v\gtrsim1.5$. Since the dominant vibration rate of a rotating component
is usually known from its nominal speed, $W$ can be selected accordingly.
The matched-window curve in Fig.~\ref{fig:rate} applies this rule, using
$W=240$ at $f_v=0.01$ and $W=80$ at $f_v=0.03$. The assisted scheme then
exceeds the conventional loop across the whole sweep, gaining $1.7$~dB at
$f_v=0.01$ instead of falling $1.8$~dB below it, and the gain stays between
$7.9$ and $10.1$~dB from $f_v=0.03$ to $0.2$. Sizing the window does not
remove the decline at the fastest rates, because only one coefficient
snapshot is obtained per adaptation period and $f_v=0.4$ approaches the
snapshot Nyquist rate of $0.5$ cycles per period.

The non-causal performance bound in Fig.~\ref{fig:rate} is nearly flat at
$38.2$~dB across the sweep, because it applies the true coefficient
trajectory and is limited only by the basis projection error and the noise
floor. The gain available to prediction therefore widens as the channel
varies faster, from $12.7$~dB at $f_v=0.05$ to $16.5$~dB at $f_v=0.2$. The
assisted scheme recovers $9.4$ and $9.9$~dB of these, or $74\%$ and
$60\%$ of the gain that perfect knowledge of the SI-channel variation
would provide.

\begin{figure}[!tb]\centering
\includegraphics[width=0.80\linewidth]{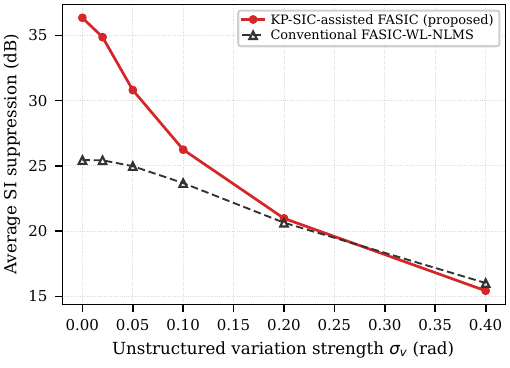}
\caption{Suppression under increasing unstructured SI variation at
$f_v=0.05$.}
\vspace{-3mm}
\label{fig:unstructured}
\end{figure}

\subsection{Unstructured variation}
We finally evaluate the sensitivity of KP-SIC to unstructured SI-channel
variation, where the coefficient motion is no longer dominated by the
quasi-periodic components modeled by the Koopman forecast.
Fig.~\ref{fig:unstructured} shows the SI suppression at $f_v=0.05$ as the
strength of the unstructured component increases. When $\sigma_v=0$, the
assisted loop improves suppression by $10.9$~dB over conventional
FASIC-WL-NLMS. As $\sigma_v$ increases, this advantage decreases because
the future coefficient variation becomes less predictable from past
snapshots. At $\sigma_v=0.2$~rad, the gain reduces to $0.3$~dB, and at the
strongest variation the assisted loop falls $0.6$~dB below the conventional
loop.

This result shows the expected tradeoff of a predictive scheme. KP-SIC is
most effective when the SI-channel variation contains structured motion
that can be identified from the snapshot history. When the unstructured
component becomes dominant, the Koopman forecast no longer provides a
reliable prediction of the next adaptation period. However, the performance
loss remains small in the tested range because the WL-NLMS correction
absorbs most of the forecast error within each adaptation period. Thus, the
assisted loop naturally approaches conventional adaptive operation as the
structured-motion assumption becomes less accurate.

\section{Conclusion}
This paper proposed KP-SIC for predictive coefficient tracking in
vehicle-mounted monostatic ISAC systems. By exploiting the structured
evolution of vibration-induced SI-channel variation, KP-SIC predicts the
cancellation coefficients for the next adaptation period and reduces the
tracking burden on the adaptive filter. In a FASIC realization with
WL-NLMS correction, the proposed scheme improves SI suppression by
$6.3$--$10.2$~dB over conventional FASIC-WL-NLMS and recovers
$60$--$74\%$ of the prediction gain available to a non-causal bound. The
results show that the observation window should span sufficient vibration
cycles to identify the dominant modes, which provides a practical
window-sizing guideline for vibration-aware SIC. Future work will address imperfect auxiliary transmit chains, real-time Koopman implementation, validation with measured platform vibration data, and extension to wideband ISAC waveforms.

\end{document}